\documentclass[pra,showpacs,twocolumn,longbibliography,secnumarabic,superscriptaddress]{revtex4-2}

\usepackage{graphicx}
\usepackage{amssymb}
\usepackage{float}
\usepackage[T1]{fontenc}
\usepackage{amsmath}
\usepackage{mathtools}
\usepackage{url}
\usepackage{hyperref}
\usepackage[dvipsnames]{xcolor}

\usepackage{amssymb}
\usepackage{dcolumn}
\usepackage{bm}
\usepackage{soul}

\begin{document}

\preprint{APS/123-QED}

\title{Electrical-control of third-order nonlinearity via Fano interference}

\author{Deniz Eren Mol}
\affiliation{Institute of Nuclear Sciences, Hacettepe University, 06800 Ankara, Turkey}%
\affiliation{Department of Nanotechnology and Nanomedicine, Faculty of Science, Hacettepe University, 06800 Ankara, Turkey}%

\author{İbrahim Asrın Üzgüç}%
\affiliation{Department of Physics Engineering, Faculty of Engineering, Hacettepe University, 06800 Ankara, Turkey}%

\author{Ulaş Eyüpoğlu}%
\affiliation{Department of Physics, Faculty of Arts and Science, Middle East Technical University, 06800 Ankara, Turkey}%

\author{Kübra Atar}%
\affiliation{Department of Chemical Engineering, Faculty of Engineering, Hacettepe University, 06800 Ankara, Turkey}%
\affiliation{TUPRAS, 35000 Izmir, Turkey}

\author{Sena Taşkıran}%
\affiliation{Department of Physics Engineering, Faculty of Engineering, Hacettepe University, 06800 Ankara, Turkey}%

\author{Taner Tarik Aytas}%
\affiliation{Department of Physics, Faculty of Science, Akdeniz University, 07058 Antalya,Turkey}%

\author{Rasim Volga Ovali}%
\affiliation{Department of Physics,  Recep Tayyip Erdogan University, 53100 Rize, Turkey}%

\author{Ramazan Sahin}%
 \email{rsahin@itu.edu.tr}
\affiliation{Department of Physics, Faculty of Science, Akdeniz University, 07058 Antalya,Turkey}%

\author{Mehmet Emre Tasgin}%
 \email{metasgin@hacettepe.edu.tr}
\affiliation{Institute of Nuclear Sciences, Hacettepe University, 06800 Ankara, Turkey}%
\affiliation{Department of Nanotechnology and Nanomedicine, Faculty of Science, Hacettepe University, 06800 Ankara, Turkey}%

\date{\today}

\begin{abstract}
	Programmable photonic computers necessitate the integration of electrically-tunable compact components into the photonic devices. In the state-of-the-art photonic quantum computers~(PQCs), phase-shift and displacement gates can be implemented in an electrically-programmable way. An efficient PQC, however, necessitates also the tuning of third or higher order nonlinearity for implementing continuous-variable~(CV) gates at a shorter sequence. 
	Here, we demonstrate that such an optical component can be designed using Fano interference and Stark effect in a nonlinear nano-plasmonic system. We study the coupling of a broadband bright plasmon mode to a narrow linewidth quantum object(s), QO(s). We show that by shifting the level-spacing of the QO via Stark effect, one can continuously tune the third-order nonlinearity gate within a picosecond response time. We also present finite-difference time domain~(FDTD) simulations that take the retardation effects into account. In addition, we also show that enhancement due to Fano interference degrades if the QOs are positioned randomly as each QO introduces different phases. This reveals the importance of the spatial extent of the QO-ensemble to be employed in the experiments.
\end{abstract}

\maketitle


\section{\label{sec:level1}Introduction}

In the past decades, we have tested fundamental phenomena on quantum optics, such as entanglement and teleportation~\cite{AsavanantGeneration}, where the dimensions of the setups were on the cm-to-m scale~\cite{yukawaHighFidelity}. Now we are in the stage of miniaturizing quantum technologies into compact integrated quantum circuits~(IQCs). A useful quantum device (e.g., for computing or information processing), however, needs to be universal and fault tolerant. Continuous-variable~(CV) quantum computers~(QCs), that employ quadrature variables of the electromagnetic field, are promising for achieving this goal~\cite{LloydQuantumContUniversal} as they employ the already-developed platforms. Moreover, photonic platforms employing CVs can also possess some advantages at learning high dimensional processes~\cite{PhotonicQuantumLearning}.

Recent progresses in the measurement-based quantum computing~(MBQC), that can be integrated with GKP encoding~\cite{GKP}, appears as an ideal platform for fault-tolerant PQCs~\cite{Larsen2025}. In MBQCs, nonlinearity generation takes place out of the IQC and squeezing gates and higher-order nonlinearity gates are implemented via coupling these ancillary states to the circuit via beam-splitter measurements. This way, the noise and loss appearing due to  the nonlinearity generation does not leak  into the circuit. Moreover,  a single circuit is operated in a loop configuration. This not only allows the sequential implementations of different gates but also  makes MBQCs very compact~\cite{TakedaExperimental}.

The universal CV photonic IQCs are constituted of beam-splitters, phase-shifters, squeezers and higher-order nonlinear components~\cite{braunstein2005quantum}. The repeated infinitesimal operation of these gates enables the universal quantum computation \cite{TakedaFurusawa}. While the former two components are electrically controllable ~\cite{OnDemandPhotonic}, electrical-control over the squeezing~\cite{gunay2023demand} and nonlinear components are also necessary in order to implement universal gates at much shorter sequences~\cite{braunstein2005quantum,Caspani2017,ModerateNonlinearity}.

In this paper, we study a scheme for the electrical-control of the third-order nonlinearity to be used in CV integrated photonic quantum computers. We consider a coupled system of a metal nanoparticle~(MNP) dimer and a quantum object~(QO, can be a quantum dot~\cite{FERRY2004298} as well as collection of ion-implanted defect-centers~\cite{NVcenters, Gcenters, denseNVcenter, EnhancedPL}). Level-spacing of the QO(s) is chosen around the frequency of the third-order nonlinear response, i.e., $\Omega_{\rm \scriptscriptstyle QO}\sim 3\omega$. The presence of the QO introduces a transparency~(Fano resonances~\cite{Stockman2010, FanoInNano, limonov2017fano, tasgin2018fano}) in the third-order nonlinear response. That is, it turns off the nonlinearity when the level-spacing is chosen as  $\Omega_{\rm \scriptscriptstyle QO}= 3\omega$ (see \autoref{fig2}-a.  On the contrary, for $\Omega_{\rm \scriptscriptstyle QO}= 2.99\omega $, Fano interference can enhance nonlinearity (see \autoref{fig2}-b).

Fano interference~(resonance) is the plasmon-analog of electromagnetically-induced transparency~(EIT), a phenomenon that has been studied at least for four decades on atomic vapors~\cite{EITimamoglu}. Presence of a QO~(such as quantum dots \cite{SemiQD,ColloidalQD}, organic molecules~\cite{OrganicMolecules}) and defect-centers (DCs)~\cite{NVcenters, Gcenters, denseNVcenter}) introduces two alternative paths for the absorption and emission of the incident light. The two paths interfere destructively, canceling the absorption~\cite{Garrido_Alzar_2002} . This path interference yields a transparency in the spectrum at the level-spacing of the quantum object~(QO)~\cite{turkpence2014engineering,limonov2017fano,tasgin2018fano}. A similar transparency appears also in the nonlinear response~\cite{limonov2017fano,tasgin2018fano}, when the QO level-spacing is chosen around the converted nonlinear frequency. 

When the level-spacing of the QO changes, i.e., by applying a voltage~\cite{shibata2013large,larocque2024tunable,miller1985electric}, analytical studies show that the nonlinear response can be altered by orders of magnitude~\cite{gunay2023demand,ullah2023electricallyprogrammablefrequencycombcompact}. This is supposed to enable a compact and continuous electrical tuning of the nonlinear response and the entanglement, thus poses a crucial component in integrated quantum circuits. As universal photonic quantum computing necessitates a third-order or higher-order nonlinearity~\cite{braunstein2005quantum}, besides the single-mode squeezing gate, here we study the Fano control of the third harmonic generation~(THG). Exact solutions of Maxwell equations show that by altering the QO level-spacing we can continuously tune the THG intensity between $10^{-1} I_0$ and $10^2 I_0$, where $I_0$ is the intensity without the presence of the QO.  These suppression and enhancement factors multiply the already appearing enhancement due to field-localization. We also study the Fano interference caused by multiple QOs that are located at different positions on the MNP. We show for the first time that they degrade the Fano enhancement effect.

We demonstrate such a control both (i) analytically, treating the plasmon modes as harmonic oscillators and the QO as two-level atoms~\cite{Scully_Zubairy_1997,premaratne2017theory,tasgin2018fano,turkpence2014engineering}, and via (ii) exact solutions of the 3D Maxwell equations using finite difference time domain~(FDTD) package of Lumerical~\footnote{See, Ansys Lumerical Inc. FDTD page}. The latter takes  the retardation effects also into account. In FDTD simulations, we model the collection of defect-centers by a Lorentzian dielectric function~\cite{Wu:10, Leng2018, Li2021} and use the experimental dielectric function for gold MNP. The third order response, $\chi^{(3)}$, is modeled exactly in FDTD which can either origin from the self-nonlinear-response of the MNP~\cite{Ginzburg2015} \cite{tepe_2024} or from molecules possessing a stronger nonlinearity \cite{Aouani2014}. 

Such a scheme should be considered as a compact device that prepares the auxiliary nonlinear pulses for the MBQC. The auxiliary pulses is mixed into the fragile quantum circuits through a beam-splitter in a measurement-based scheme~\cite{QCoverCV} for implementing the third-order-nonlinearity gate~\cite{CVQC}. As system employs plasmonic excitations, response time is about picosecond that is much faster than the state-of-the-art MBQC~\cite{TDM_MBQC} where the beam-splitter is tuned electrically. Actually, the integration of this scheme into MBQC is also limited by the measurement time in the MBQC. Moreover, as the level-spacing of the QO is at the nonlinear frequencies, the linear response is not modified from pulse to pulse.

The paper is organized as follows. We first study an analytical model on which we demonstrate why and how Fano interference~(also referred as resonances) takes place on the third-order nonlinear response. Second, we also take the retardation effects into account by performing FDTD simulation. We demonstrate that third-order nonlinearity, third harmonic generation in our case, can either be  suppressed or enhanced depending on the choice of the QO level-spacing. We propose a method to electrically-tune the THG intensity by shifting the level-spacing. Next, we show that the enhancement obtained by a QO positioned at the hot spot, can degrade if one places other QOs at different positions of the MNP surface. This takes place because the two Fano interference possess two different phases that originate from the differentiated positions of the different QOs.  This degrades the overall enhancement for increasing number of randomly positioned QOs. The last section contains our conclusions and outlook.

\section{Fano resonances in the third-order nonlinear response}

 Here, we present a scheme for realizing an electrically-programmable third-order nonlinear component that also satisfies the aforementioned miniaturization requirements. We choose plasmonics as a platform because (i) it occupies a small space (compactness) ~\cite{ozbay2006plasmonics}, (ii) plasmonic localization effect enhances the nonlinearity~\cite{wu2010quantum}, and (iii) it possesses the small response time required for fast operations.
 We first demonstrate the effect analytically. On a single equation, Eq.~(\ref{eq18}), we demonstrate why such a suppression and enhancement in the THG appear for different choices of the QO level-spacing. Then, in the next section, we demonstrate the phenomenon via exact solutions of the 3D Maxwell equations.

Plasmonic nanostructures~(PNSs) localize the incident electromagnetic field into nm-size hotspots as surface plasmon oscillations. This near-field is enhanced several orders of magnitude~\cite{HentschelQuantitative}. Noting that the THG is proportional to the third power of the linear enhancement, the nonlinear frequency conversion is also enhanced orders of magnitude. Moreover, if the generated high-energy mode is also localized, the enhancement increases even more. (See the overlap integrals below.) 

In this paper, we consider a dimer of two spherical gold nanoparticles. The hotspot appears between the two dimers. This is because the dimer is pumped by a laser of frequency $\omega$ that is polarized along the axis of the dimer. 
We position a quantum object ---can be a quantum dot, defect center \cite{chu2015quantumopticsnitrogenvacancycenters} or molecular aggregates~\cite{AnomalousAggregates}--- at the hotspot of the plasmonic nanostructure~(PNS). We choose the level-spacing of the QO around the third harmonic frequency, i.e,  $\Omega_{\rm \scriptscriptstyle QO}\sim 3\omega$. The QO residing at the hotspot interacts very strongly with the high-energy plasmon mode whose resonance is around $\Omega_3\sim 3\omega$. This interaction introduces the Fano interference at nonlinear conversion without altering its linear response. The lower-energy plasmon mode of resonance $\Omega_1$ is pumped by the laser. Three of the $\omega$-frequency plasmons in the $\Omega_1$ mode combine to generate a single third harmonic plasmon $3\omega$ in the $\Omega_3$ mode. Experiments demonstrate that nonlinear processes take place over plasmons~\cite{grosse2012nonlinear} on PNSs. This is because overlap intergrals for such processes becomes very large owing to the localization~\cite{tasgin2018fano,Ginzburg2015}.  Both the incident~(fundamental) field ~($\omega$) and the converted field ~($3\omega$) are localized, yielding a superior overlap integral ~\cite{tasgin2018fano,Ginzburg2015}. 

We change the level-spacing of the QO~($\Omega_{\rm \scriptscriptstyle QO}$) and investigate the generated third harmonic~(TH) signal. In practice, one can shift the QO level-spacing by applying a voltage on it~\cite{shibata2013large,larocque2024tunable,miller1985electric}. Third harmonic generation of the PNS can either be suppressed for $\omega_{\rm \scriptscriptstyle QO} = 3\omega$ or be enhanced for $\Omega_{\rm \scriptscriptstyle QE} \simeq 2.99\omega$.  Tuning the applied voltage between $V$=0--1V, one can tune~(modulate) the intensity of THG between $10^{-1}$--$10^2$ in such a device.  We simulate the proposed plasmonic system by exact solutions of the nonlinear 3D Maxwell equations. We use finite-difference time-domain~(FDTD) module of the Lumerical~\footnote{Lumerical Inc. FDTD}.

\begin{figure}
    \centering
    \includegraphics[width=\linewidth]{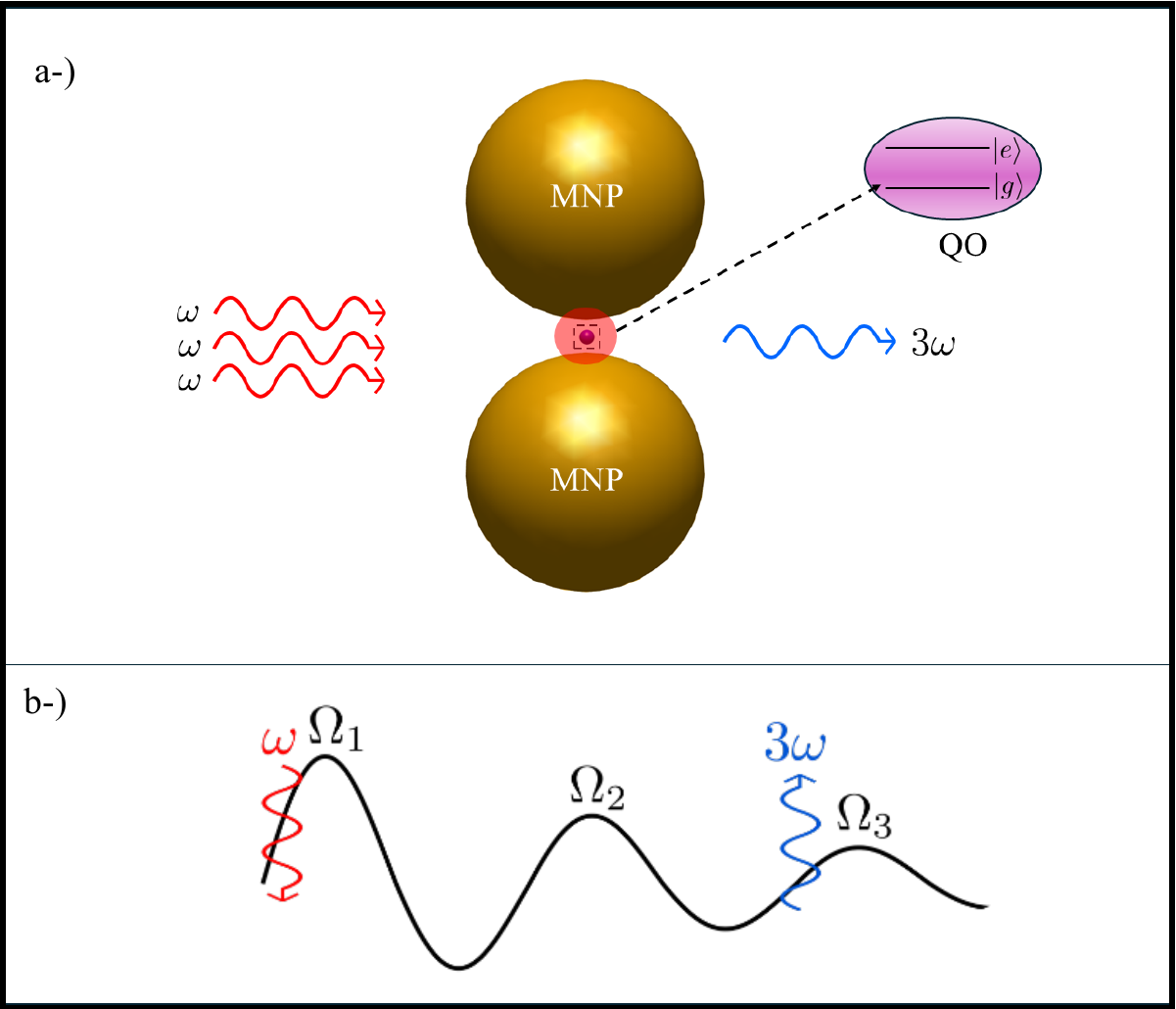}
    \caption { (a) Control of third harmonic generation signal by a quantum object~(QO) residing at the hotspot of a metal nanoparticle~(MNP) dimer. Pump photons of frequency $\omega$ are localized at the hotspot as plasmon oscillations takes place in the $\Omega_1$-mode. Localization gives rise to several orders of magnitude enhancement in the third harmonic generation~(THG).  The level-spacing of the QO is chosen around the THG frequency, $\Omega_{\rm \scriptscriptstyle QO}\sim 3\omega$. Level-spacing  controls the nonlinear signal.    $\Omega_{\rm \scriptscriptstyle QO}= 3\omega$ turns off the THG while  $\Omega_{\rm \scriptscriptstyle QO}\simeq 2.99\omega$ enhances the THG. Applied voltage shifts the level-spacing~\cite{shibata2013large,larocque2024tunable,miller1985electric}. (b) The plasmon modes and corresponding fundamental and third harmonic frequencies.}
\label{fig1}
\end{figure}

In this section, we study the system depicted in Fig.~\ref{fig1} analytically. We first derive a single equation to explore the enhancement/suppression mechanism. On this single equation, we demonstrate why Fano-suppression and Fano-enhancement effects appear. In the next section, we will also perform the FDTD simulations and compare our analytical model with the exact solutions of the Maxwell equations.

\subsection{Analytical Model}

The plasmon modes related with the THG process are depicted in Fig.~\ref{fig1}b. The incident photons of frequency $\omega$ pumps the $\hat{a}_1$ plasmon mode of resonance $\Omega_1$. Such a process is governed by the hamiltonian
\begin{equation}
	H_{\rm p}= \left( E_0 \int d^3{\bf r} \:  u_1({\bf r})   e^{ikz}  \right) \hat{a}_1 e^{-i\omega t} + H.c. 
	\label{eq2}
\end{equation}
whose strength is determined by the overlap integral (that is equal to $\varepsilon_p$) inside the parenthesis. Here, $u_1({\bf r})$ is the spatial profile of the $\hat{a}_1$ plasmon mode which is localized between the two MNPs, the the hotspot.  The hamiltonian can be rewritten as
\begin{equation}  
	\hat{H}_{\rm p}=i\hbar(\varepsilon_p{\hat{a}_1}^\dag e^{-i\omega t}-\varepsilon_p{\hat{a}_1}e^{i\omega t}) . 
	\label{eq5}
\end{equation} 
The intense plasmon near-field at the hotspot is highly enhanced due to the field localization. Then, nonlinear processes comes into the play. Three $\omega$-frequency plasmons in the $\Omega_1$~($\hat{a}_1$) mode combine to generate a $3\omega$-frequency plasmon  in the $\Omega_3$~($\hat{a}_3$) mode. In PNSs, such frequency conversion phenomena are shown to take place over the plasmon modes~\cite{grosse2012nonlinear} because of the large overlap integrals~\cite{tasgin2018fano,Ginzburg2015}. That is, the THG process is governed by the hamiltonian
\begin{equation}
	H_{\rm \scriptscriptstyle THG}= \left(\int d^3 \textbf{r} \: u_3^*(\textbf{r}) u_1^3 \textbf{(r)} \chi^{(3)} (\textbf{r}) \right)\hat a_3^\dagger\hat a_1 \hat a_1 \hat a_1 +H.c.,
	\label{eq3}
\end{equation}
whose dynamics is determined by the overlap integral in the parenthesis that we set equal to $\hbar \chi_{3}$. The hamiltonian looks like
\begin{equation}  
	\hat{H}_{\rm \scriptscriptstyle THG}=\hbar\chi_3({\hat{a}_3}^\dagger{\hat{a}_1}{\hat{a}_1}{\hat{a}_1}+{\hat{a}_1}^\dagger{\hat{a}_1}^\dagger{\hat{a}_1}^\dagger{\hat{a}_3}). 
	\label{eq6}
\end{equation}
Above, $u_3({\bf r})$ is the spatial profile of the high-energy $\hat{a}_3$ plasmon mode of resonance $\Omega_3$ in which $3\omega$ plasmon emerges. $\chi^{(3)} (\textbf{r})$ is the third-order nonlinear response of the material. The THG can origin both from the MNPs themselves~\cite{kauranen2012nonlinear,Ginzburg2015,tepe_2024} or from molecules ~\cite{ren2014enhanced}. $\chi^{(3)} (\textbf{r})$ refers to the third-order nonlinear susceptibility of the material and its spatial profile. Thus, the overlap integral becomes very large if the hotspot profiles $u_1({\bf r})$ and $u_3({\bf r})$ overlap with each other and the nonlinear material $\chi^{(3)} (\textbf{r})$.
We should note that the well known selection rules~\cite{chen2014symmetry} follow from the such overlap integrals~\cite{tasgin2018fano}. 

The control of the THG is achieved via the QO that is coupled to the $\Omega_3$ plasmon mode. The interaction is governed by the hamiltonian
\begin{equation}  
\hat{H}_{\rm int}=\hbar f \:  {\hat{a}_3}^\dag \: |g\rangle\langle e|+ H.c., 
	 \label{eq9}
\end{equation}
where $f$ is the coupling strength of the QO to the $\Omega_3$-mode. $|g\rangle$,  $|e\rangle $ are the ground and excited states of the QO. 

Including also the energies of the plasmons
\begin{equation}  
	{\hat{H}_{\rm \scriptscriptstyle MNP}}=\hbar\Omega_1{\hat{a}_1}^\dag{\hat{a}_1}+\hbar\Omega_3{\hat{a}_3}^\dag{\hat{a}_3}
	\label{eq7}
\end{equation}
and the QO 
\begin{equation}  
	\hat{H}_{\rm \scriptscriptstyle QO}=\hbar\omega_{eg}|e\rangle\langle e| \equiv \Omega_{\rm \scriptscriptstyle QO} |e\rangle\langle e| ,  
	\label{eq8}
\end{equation}
the total hamiltonian of the system can be written as
\begin{equation}
	\hat H_{\rm total}= \hat H_{\rm \scriptscriptstyle MNP}+\hat H_{\rm \scriptscriptstyle QO} +  \hat H_{\rm p} + \hat H_{\rm \scriptscriptstyle THG} + \hat H_{\rm int}.   
	\label{eq4}
\end{equation}
Here, $\omega_{eg}\equiv \Omega_{\rm \scriptscriptstyle QO}$ is the level-spacing of the QO, i.e., energy difference between the excited and the ground states.

\subsection{\label{sec:level2} Langevin Equations and Time Evolution}

We can determine the time evolution of the operators using the Heisenberg equations of motion, e.g., $i\hbar \dot {\hat a}_i=[\hat a_i,\hat H])$. In this study, we are interested in only the field amplitudes ---i.e., not interested in quantum optics features such as entanglement~\cite{gunay2023demand,ullah2023electricallyprogrammablefrequencycombcompact}. Thus, we replace the operators with the field amplitudes $\hat{a}_i \to \alpha_i$~\cite{premaratne2017theory,turkpence2014engineering}. Thus, the equation of motion for the field amplitudes become
\begin{eqnarray}  
&& \dot{\alpha}_1=-(i\Omega_1+\gamma_1)\alpha_1-i3\chi_3 \alpha_1^{*3}\alpha_3 +\varepsilon_\rho e^{-i\omega t},  
\label{eq10}
\\
&& \dot{\alpha}_3=-(i\Omega_3+\gamma_3)\alpha_3-i\chi_3\alpha_1^{3}-if \rho_{ge}, 
\label{eq11}
\\
&& \dot{\rho}_{ge}=-(i\Omega_{\rm \scriptscriptstyle QO}+\gamma_{eg})\rho_{ge}+i f\alpha_3 (\rho_{ee}-\rho_{gg}),  
\label{eq12}
\\
&& \dot{\rho}_{ee}=-\gamma_{ee}\rho_{ee}+i\{ f \alpha_3^*\rho_{ge}- f \alpha_3 \rho_{eg}\}.  
\label{eq13}
\end{eqnarray}
Here, $\gamma_1$, $\gamma_3$ are the damping rates of the  plasmon modes and $\gamma_{ee}$, $\gamma_{eg}$ are the diagonal and off-diagonal decay rates for the QO~\cite{Scully_Zubairy_1997}. For a single QO,  $\gamma_{ge}$=$\gamma_{ee}/2$. $\rho_{ee}$ and $\rho_{gg}$ are the diagonal density matrix elements for the QO associated with the occupation probabilities of the excited and ground states, respectively. $\rho_{eg}$ is the off-diagonal matrix element responsible for polarization~\cite{Scully_Zubairy_1997}. 

For an optical drive~(pump) of frequency $\omega$ the steady-state solutions can be determined by inserting the slowly-varying envelopes
\begin{eqnarray} 
\alpha_1(t)=\tilde{\alpha}_1e^{-i\omega t}, \quad \alpha_3(t)=\tilde{\alpha}_3e^{-i3\omega t}, 
	\\
\rho_{ee}(t)=\tilde{\rho}_{ee} \quad \text{and} \quad \rho_{ge}(t)=  \tilde{\rho}_{ge}e^{-i3\omega t}    .
\label{eq14}
\end{eqnarray}
In addition, one also uses the condition  $\rho_{ee}+\rho_{gg}=1$. Here, $y$ is the population inversion  $y=\rho_{ee}-\rho_{gg}$.

\subsection{\label{sec:level2} Why suppression and enhancement occur?}

We can develop a simple understanding on the control of the THG amplitude $\alpha_3$ by investigating the steady-state solutions of  Eq.~(\ref{eq11})
\begin{equation}
\tilde{\alpha}_3(-3i\omega)=-(i\Omega_3+\gamma_3)\tilde{\alpha}_3-i\chi_3 \tilde{\alpha}_1^3-if \tilde{\rho}_{ge}  
\label{eq16}
\end{equation}
and Eq.~(\ref{eq12})
\begin{equation}  
\tilde{\rho}_{ge}(-3i\omega)=-(i\Omega_{\rm \scriptscriptstyle QO}+\gamma_{eg})\tilde{\rho}_{ge}+(if\tilde{\alpha}_3 y) .
\label{eq17}
\end{equation}
A simple algebra gives
\begin{equation} 
\tilde{\alpha}_3= \frac{ -i\chi_3 } { [i(\Omega_3-3\omega)+\gamma_3] - \frac{|f|^2 y}{i(\Omega_{\rm \scriptscriptstyle QO}-3\omega)+\gamma_{eg}} }\tilde{\alpha}_1^3  
\label{eq18}
\end{equation}
for the steady-state third harmonic mode amplitude $\tilde{\alpha}_3$ in terms of the amplitude of the fundamental mode  $\tilde{\alpha}_1$ and the level-spacing of the QO, $\Omega_{\rm \scriptscriptstyle QO}$. 

We scale all frequencies by the pump frequency $\omega$, i.e., $\omega=1$. The typical values of the damping rates in these units are $\gamma_{ee}\sim 10^{-5}$, $\gamma_1\sim \gamma_3  \sim 0.1$, and $f=0.1$~\cite{wu2010quantum,Leng2018,tasgin2018fano}.  

One can demonstrate the  suppression~(switch off) and enhancement~(switch on) phenomena on this single equation. When one tunes the level-spacing to $\Omega_{\rm \scriptscriptstyle QO}=3\omega$, the second term in the denominator becomes $y\:|f|^2/\gamma_{eg}$. In the scaled units, this term becomes very large compared to the first term of the denominator. Thus, the THG is vanishingly suppressed and there appears a transparency window in the third harmonic response as depicted in Fig.~2a. On the contrary, one can also choose~(tune) the $\Omega_{\rm \scriptscriptstyle QO}$ such that the imaginary part of the second term in the denominator cancels the non-resonant term ($\Omega_3-3\omega$). Then, the third harmonic conversion is brought into resonance. Thus, it is enhanced as one can see in Fig.~\ref{fig2}b.  

Therefore, tuning  $\Omega_{\rm \scriptscriptstyle QO}$ between the suppression and the enhancement values, one can continuously tune the THG signal. This is depicted in Fig.~\ref{fig3}. 
In the next section, we present these phenomena also via exact solutions of the 3D nonlinear Maxwell equations.

\section{Exact solutions of the 3D nonlinear Maxwell equations }
\label{chap:foo}

We also perform finite difference time domain simulations for the nonlinear 3D Maxwell equations. We use the well-known toolbox Lumerical~\footnote{See Lumerical Inc. Material modelling page}. Two gold metal nanoparticles~(MNPs) of diameter 100 nm are located along the z-axis with a gap size of 10 nm. See Fig.~\ref{fig1}. We use the experimental dielectric functions for the MNPs \cite{OptConstOfNobleMetals}. We also implement a constant nonlinear susceptibility $\chi^{(3)}=2.45 \times 10^{-19}$ ${\rm m^2}/{\rm V}^2$~\cite{boyd2014third}  on the MNP. Such a nonlinearity  can originate from the self nonlinear response of the MNPs~\cite{kauranen2012nonlinear,Ginzburg2015,tepe_2024} or from high nonlinear-response molecules residing on the MNP surface ~\cite{ren2014enhanced}. 

We position a QO with a narrow linewidth of $\gamma_{eg}=10^{10}$ Hz. We take the radius of the QO as 3 nm. We vary the level-spacing of the QO between 1693.7 THz and 1696 THz for controlling the produced THG signal. See Fig.~\ref{fig3}. Such a tuning can be performed by applying a potential difference on the QO~\cite{shibata2013large,larocque2024tunable,miller1985electric}. We model/simulate the QO by a Lorentzian dielectric function~\cite{wu2010quantum,Leng2018,Li2021}
\begin{equation}
	\epsilon(\omega)=\epsilon_{\rm b}+ f_{\rm osc} \frac{ \Omega_{\rm \scriptscriptstyle QO}^2}{\Omega_{\rm \scriptscriptstyle QO}^2-2i \gamma_{eg} \omega - \omega^2},
	\label{eq19}
\end{equation}
where $\Omega_{\rm \scriptscriptstyle QO}$ is the resonance~(level-spacing), $\gamma_{eg}=10^{10}$ Hz is the linewidth, $f_{\rm osc}=0.4$~\cite{leistikow2009size, opticalstarkmetrology} is the dimensionless oscillator strength~\cite{wu2010quantum,Leng2018,Li2021} and $\epsilon_{\rm b}=1.5$ is the background susceptibility. 
%

%

\begin{figure}
    \centering
    \includegraphics[width=\linewidth]{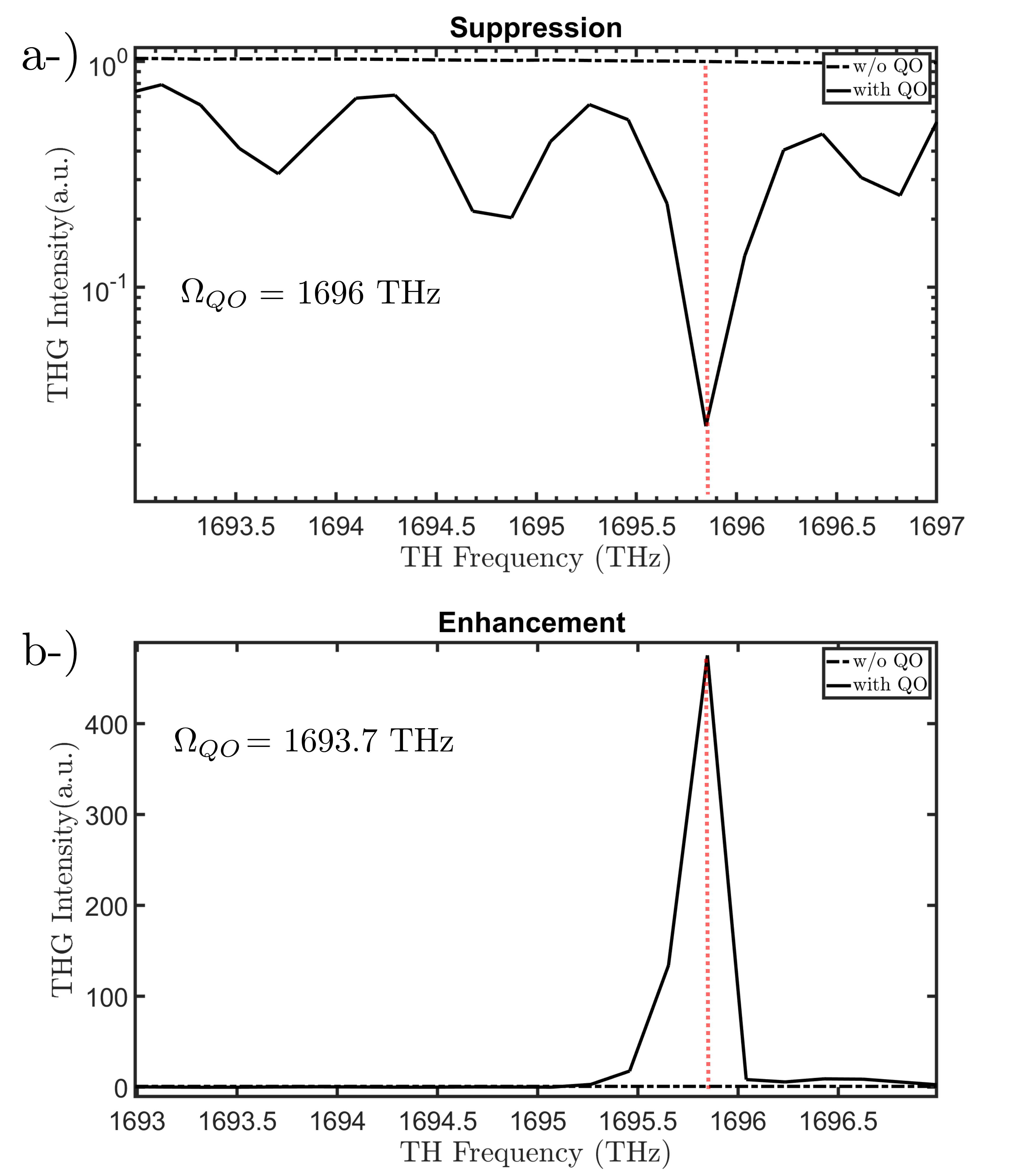}
    \caption{Demonstration of the two opposite phenomena, (a) suppression and (b) enhancement of the third harmonic signal, by the exact solutions of the 3D nonlinear Maxwell equations. For two different choices of the quantum object level-spacing (a) $\Omega_{\rm \scriptscriptstyle QO}=1696$ THz and (b) $\Omega_{\rm \scriptscriptstyle QO}=1694$ THz, suppression and enhancement are observed at the same pump frequency $3\omega=1695.8$ THz. }
\label{fig2}
\end{figure}

In Fig.~\ref{fig2}a, we set $\Omega_{\rm \scriptscriptstyle QO}=1696$ THz. We observe that the THG intensity is suppressed down to $10^{-1}$ times of the THG intensity calculated without the presence of the QO. In Fig.~\ref{fig2}b, we set~(shift) the level-spacing of the QO to $\Omega_{\rm \scriptscriptstyle QO}=$1693.7 THz and observe that the THG signal is enhanced to about 400 times of the original~(without QO) THG signal. Both suppression and enhancement take place at the same operation frequency $3\omega=1695.8$ THz.

In Fig.~3, we demonstrate how the proposed device should work when $\Omega_{\rm \scriptscriptstyle QO}$ is tuned. 
We fix the operation frequency of the device to  $3\omega=1695.8$ THz and vary the level-spacing of the QO between $\Omega_{\rm \scriptscriptstyle QO}=$1693.7 THz and 1696 THz. Such a tuning can be performed using the Stark effect~\cite{shibata2013large,larocque2024tunable,miller1985electric, StarkControlOfPlexcitonic} by applying a potential difference less than 1 Volt. We observe that the THG intensity can be continuously modulated by 3 orders of magnitude.

\begin{figure}
    \centering
    \includegraphics[width=\linewidth]{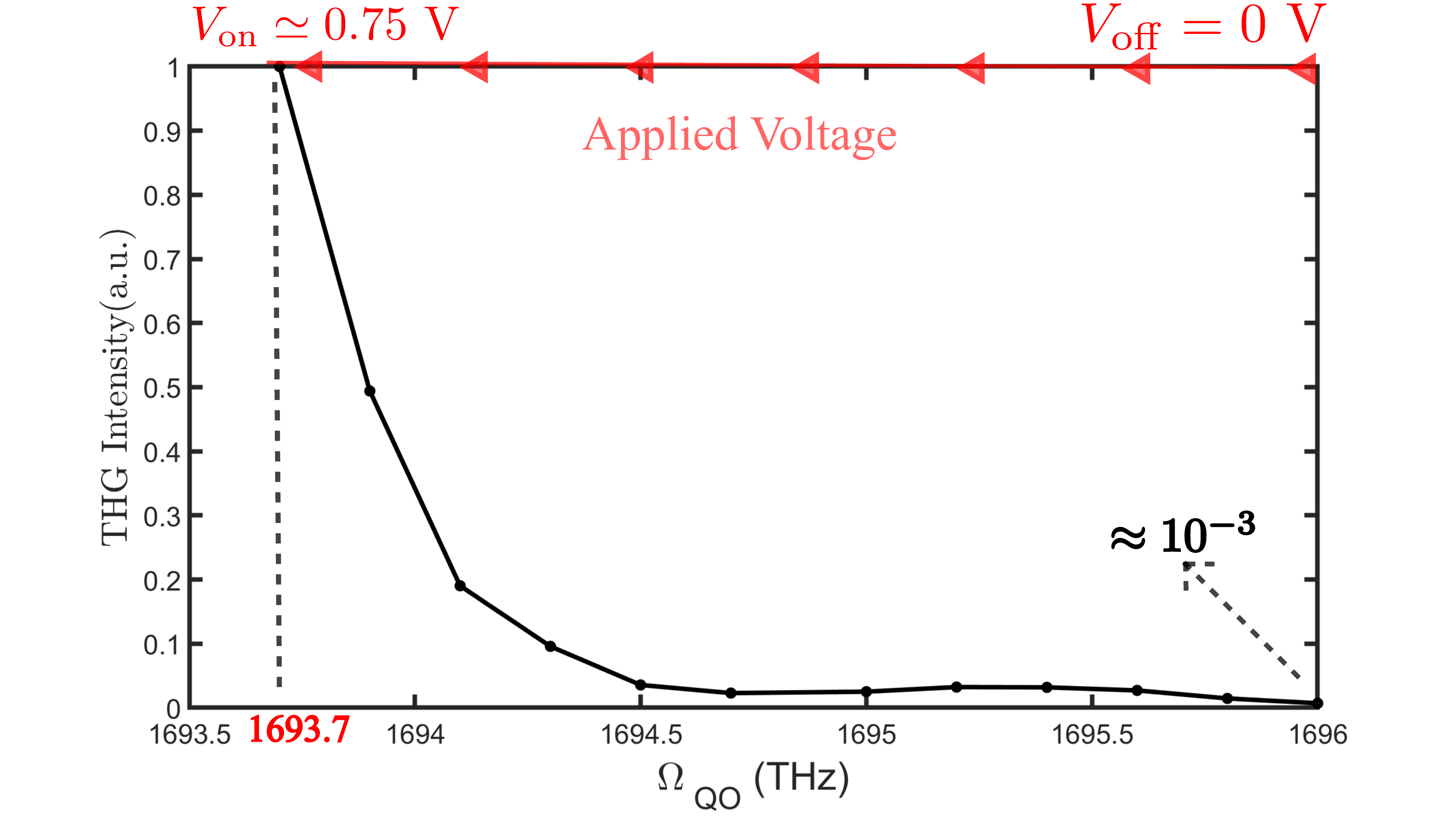}
    \caption{Continuous electrical tuning of the third harmonic intensity by shifting the level-spacing of the QO. Different level-spacing results in different path interference effects that can be explainv ia the denominator of Eq.~(\ref{eq18}). Such a tuning can be performed by applying a potential-difference on the QO~\cite{shibata2013large,larocque2024tunable,miller1985electric} that shifts the level-spacing. The required voltage ---calculated crudely from the $\Delta \Omega_{\rm \scriptscriptstyle QE}$-$V$ graph presented in the experiments~\cite{shibata2013large,larocque2024tunable,miller1985electric}--- is less than 1 Volt. The presented data is the exact solution of the 3D nonlinear Maxwell equations for the system depicted in Fig.~\ref{fig1}a.
    }
\label{fig3}
\end{figure}

 \begin{figure}
    \centering
    \includegraphics[width=\linewidth]{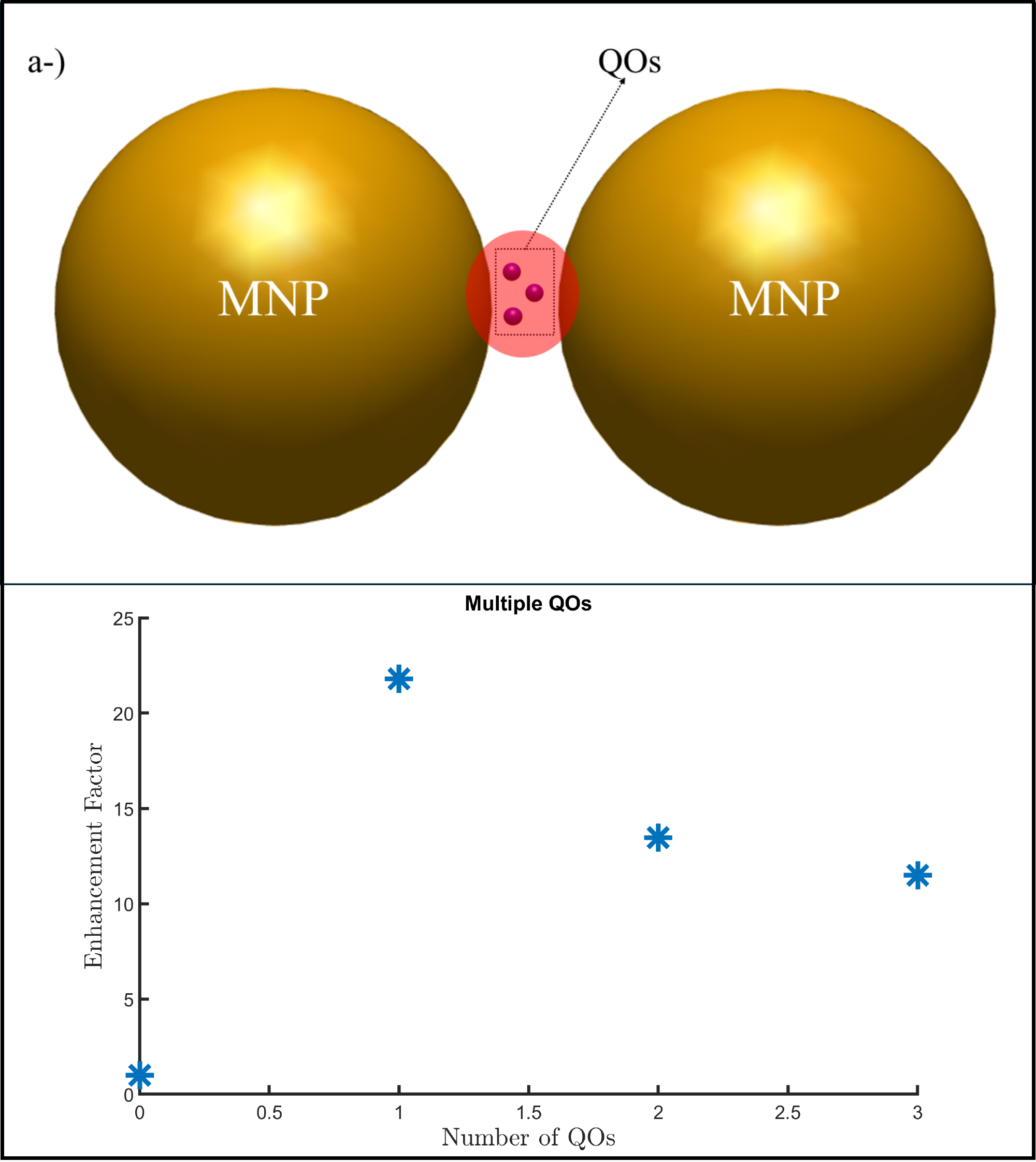}
\caption{(a) Fano enhancement of the THG when more than one QOs are placed in the gap at different positioned. (b) Because of the retardation effect, QOs located at different positions take different phases. There is still a very strong Fano enhancement but it degrades by a factor of 1/2 with respect to a single QO case. 
    	Exact solutions of the 3D nonlinear Maxwell equations. }
\label{fig4}
\end{figure}

\section{Fano Interference with more than one QOs}

The literature studies Fano interference effect (usually the linear one) caused by a single QO, a group of QOs sticking together or Fano resonance appearing due to a dark mode~\cite{wu2010quantum,Leng2018,Li2021,limonov2017fano,zhang2013ultrafast}.  
In this section, we  inquire what happens if the path interference takes place due to more than one QOs that are located at different positions on a metal nanostructure. In order to investigate this case, we place several QOs of the same size at different positions near the MNP-dimer. See Fig.~\ref{fig4}a. In Fig.~\ref{fig4}b, we plot the THG enhancement factors for different number of QOs. We observe that Fano-enhancement is maximum for a single QO. For two or more QO sitting at different positions, the enhancement is still present with a good ratio but starts to decrease compared to using only a single QO. 

Such a phenomenon takes place due to the retardation effects. Different QOs introduce terms with different phases in the denominator of $\tilde{\alpha}_3$ in Eq.~(\ref{eq18}). Thus, the cancellation of the non-resonance ($\Omega_3-3\omega$) term degrades. However, the enhancement factor is still large. Actually, we observed such an effect in our experimental on the stochastic Fano-enhancement of the surface enhanced Raman scattering~(SERS)~\cite{demirtas2025qupersquantumenhancedraman}.  QOs sitting at random position on the metal nanostructures degrade the Fano enhancement factor when ratio for the number of QOs to the number of MNSs exceeds unity.

\section{Conclusions and Outlook}

We show that third order nonlinear response of a plasmonic system can be continuously tuned using the Fano interference phenomenon~\cite{wu2010quantum,Leng2018,tasgin2018fano}. We demonstrate this using the exact solutions of the 3D nonlinear Maxwell equations. The presence of a (or a collection of) QO(s) introduces a transparency in the third order response of a MNP system. By shifting the level-spacing of the QO(s), one can tune the position of the transparency. 
So, one can control the third order nonlinear response.  A 1 volt potential difference suffices to perform this tuning.

We demonstrate the phenomenon both using a basic analytical model and solving the 3D nonlinear Maxwell equations exactly. The latter also takes the retardation effects into account. This scheme can be employed for achieving a fast~(picosecond~\cite{zhang2013ultrafast}) control over nonclassicality of the auxiliary pulses generated to perform measurement-based quantum computing~(MBQC). In MBQC, fragile quantum states are not inserted into the noisy nonlinearity-generating devices. The auxiliary nonlinearity-possessing pulses are generated outside the circuit, then they are coupled with the fragile quantum states through  a beam splitter gate. As an other advantage of our scheme, the linear response of the generated pulses is not modified. This is because, the level-spacing of the QO is at nonlinear frequencies.

We also show that Fano enhancement can degrade if different ensembles of defect-centers are located at different positions on the surface of the MNP-dimer. This takes place because of the different phase factors appearing due to the differentiated overlap integrals. Actually, this effect is observed in a recent experiment~\cite{demirtas2025qupersquantumenhancedraman} on the Fano-enhancement of surface enhanced Raman scattering~(SERS).

\begin{acknowledgments}
DEM, TTA, RS and MET  acknowledge support from TUBITAK-Project No. 1001-123F156. RS and MET acknowledge support from TUBITAK-Project No. 3501-121F030 and TUBITAK-Project No. 1001-119F101. RVO and MET acknowledge support from TUBITAK 1001-121F141.
\end{acknowledgments}

\bibliography{reference}

\end{document}